\documentstyle[numreferences]{crckapb} 

\begin{opening}
\title{NONABELIAN DEBYE SCREENING, TSUNAMI WAVES, AND WORLDLINE FERMIONS}
\author{ROBERT D. PISARSKI}
\institute{Department of Physics\\
           Brookhaven National Laboratory\\
	   Upton, N.Y.  11973  U.S.A.}
\end{opening}
\runningtitle{NONABELIAN DEBYE SCREENING}
\begin{document}
\begin{abstract}
I give a pedagogical review of the derivation
for the effective lagrangian for nonabelian
Debye screening, or hard thermal loops.
Following Kelly, Liu, Lucchesi, and Manuel, 
I give the simplest derivation possible,
using classical kinetic theory.

The result is valid not just for a thermal,
but for an arbitrary initial distributions.
I use this to study the evolution, at short
times, of a gluonic ``tsunami wave''.

I also suggest how classical kinetic theory
may arise at one loop order.
Using the wordline representation of the one loop effective action, 
I follow D'Hoker and Gagn\'{e} to replace the Wilson line
by an integral over worldline fermions.  A bilinear of
these worldline fermions naturally defines a nonabelian
charge, whose equation of motion is Wong's equation.

\end{abstract}
\section{Introduction}

In this talk I review our undertanding of plasmas
at very high temperatures \cite{htl}-\cite{lebellac}.  
While most of the detailed
discussion is phrased in terms of a plasma near
equilibrium at a temperature $T$, I show that
the essential result holds not just for a thermal,
but for arbitrary, initial distributions of particles.
The crucial assumption is that the initial density
of particles is large.  For a thermal distribution,
this means that the temperature is very high,
much larger than any mass
scale, $T \gg m$.  

There are two major applications of this work.
The first is to the behavior of $QCD$ in a ``deconfined''
phase.  This is relevant to the central region of
heavy ion collisions at high energies, as has been
explored at the $SPS$, and will be studied at
$RHIC$ and $LHC$.
The second is to the electroweak theory,
in a phase in which the Higgs phase evaporates. 
This has become of especial interest with regards
to baryeogenesis at the electroweak scale.

All calculations are done using perturbation
theory.  At zero temperature, the normal expansion parameter is
the fine structure constant, $\alpha = g^2/4\pi$; 
in $QCD$, if one is lucky, $\alpha$ is of order 
one.  For systems at nonzero temperature, however,
the relevant expansion parameter is not $\alpha$
but the coupling constant itself, $g$.  Thus
for $QCD$, if $\alpha \sim 1.$, $g \sim 3.$, and
clearly a perturbative approach is invalid.  
It is probably a reasonable approximation in
the electroweak theory, where $g \sim .3$.

My philosophy is that in order to understand
any theory, one must have some controlled and calculable
limit.  It certainly should be able to give us some
sort of picture of the important quasiparticle modes
and their interactions.  In a strong coupling regime
the true quasiparticles may behave differently, but
one must at least know what behavior that they must
match onto in weak coupling.

Although it was not clear at first, the basic
physical phenomenon which I will be studying is
that of Debye screening.  While it sounds
prosaic, even to
lowest nontrivial order in the coupling constant,
for nonabelian gauge theories
the physics is remarkably involved.
For static charges, the screening is unremarkable,
like that of nonrelativistic systems.
It is only when one considers the screening of moving
charges - that is, the dynamics of the system - that
a wonderfully rich structure emerges.

This is suggestive.  Numerical simulations of
lattice gauge theory have clearly demonstrated their
power to compute the behavior of gauge theories in
equilibrium.  These numerical simulations have shown
that the phase structure of $QCD$ depends upon the
number and masses of the quark flavors in a unexpectedly
intricate manner.  At present, however, lattice gauge
theory is only efficient for computing the behavior
the nature of quantities in equilibrium.  The
present discussion shows that even perturbation theory
shows that there will be many new suprises in considering
truly dynamical phenomenon.

To appreciate the Debye lagrangian one really has to apply it
in detailed calculations.  For reasons
of sloth, in these proceedings
I concentrate exclusively on how to derive the 
effective Debye lagrangian, and not to what uses it can be put.
The nonabelian Debye lagrangian
was first derived by brute force, after
a laborious and complicated analysis of diagrams in
perturbation theory \cite{htl}.
Blaizot and Iancu then derived it by a mean field
approximation to the Schwinger-Dyson equations,
which gives a kind of semiclassical kinetic theory \cite{blaizotiancu}.
My derivation herein uses classical kinetic theory, and
copies that of Kelly, Liu, Lucchesi, and Manuel \cite{kelly}.
At one point, I use a trick of Brandt, Frenkel,
and Taylor \cite{frenkel}.

I first derive the effective lagrangian in $QED$,
where the gauge principle is elementary, and then
that for $QCD$.  The beauty of classical
kinetic theory is that, after invoking
the nonabelian gauge symmetry,
the derivation is as simple as for the abelian theory.
My treatment is meant to be elementary,
appropriate for those who have never had any exposure 
to a nonabelian gauge theory.  This is only possible
because of the simplicity of classical kinetic theory.

There is another advantage of classical kinetic theory:
it appears to be valid not just for thermal, but for
arbitrary initial distributions.  In sec. 5 I use
this to study what I term a gluonic ``tsunami wave''.

The apparent peculiarity of classical kinetic theory 
is that it is necessary to introduce a nonabelian
charge, $Q$, and its attendant equation of motion, known
as Wong's equation \cite{wong}.  
Certainly to most high energy theorists, the introduction
of $Q$ appears {\it ad hoc} and a little peculiar.  
In the final section I outline a possible way in which $Q$
arises directly from the worldline formalism.

\section{Classical kinetic theory in $QED$}

The lagrangian of massless $QED$ is, of course,
\begin{equation}
{\cal L} = \bar{\psi}\gamma^\alpha D_\alpha \psi
+ \frac{1}{4} F_{\alpha \beta}^2 \, .
\end{equation}
The covariant derivative and field strength are
\begin{equation}
D_\alpha = \partial_\alpha - i e A_\alpha \; , \;
F_{\alpha \beta} = \partial_\alpha A_\beta
- \partial_\beta A_\alpha \; .
\end{equation}
This is invariant under local gauge transformations,
\begin{equation}
D_\alpha \rightarrow \Omega^\dagger D_\alpha \Omega \; , \;
\psi \rightarrow \Omega^\dagger \psi \; .
\label{e1}
\end{equation}
The transformation of the gauge field is more familiarly
written as 
\begin{equation}
\Omega = e^{i e \omega} \; , \;
A_\alpha \rightarrow A_\alpha + \partial_\alpha \omega \; .
\label{e2}
\end{equation}
For the nonabelian theory, however, it will turn out that
thinking about covariant derivatives, as in (\ref{e1}),
is much more useful than in thinking about gauge potentials,
as in (\ref{e2}).  This difference isn't apparent in the
abelian theory.

We then need the classical equations of motion.  To be logically
complete I should start from the lagrangian and 
derive the appropriate equations in the limit of high density.
A way in which to do this is sketched in sec. 6.  For now
I will take the equations of motion for granted, since they
are utterly standard: introducing 
the position, $x^\alpha$, and the momentum, $p^\alpha$,
there is just the definition of the momentum,
\begin{equation}
p^\alpha = \frac{d x^\alpha}{d \tau} \; , 
\label{e3}
\end{equation}
and the Lorentz force equation,
\begin{equation}
\frac{d p_\alpha}{d \tau} = e \; F_{\alpha \beta} \; p^\beta \; .
\label{e4}
\end{equation}
The field is assumed to be massless, so 
``$\tau$'' is not really a proper time, just an affine parameter
which labels the worldline of the particle.

As typical of classical kinetic theory, I introduce the density
of single particles, $f(x,p)$.  This is assumed to satisfy
the Boltzmann equation in the collisionless approximation,
\begin{equation}
\frac{d }{d\tau}\, f(x,p) = 0 \; .
\label{e5}
\end{equation}
Viewing $f(x,p)$ as a density in phase space, this is
a type of Liouville equation.

The only other quantity required is the current.  
In the classical approximation this is merely the 
product of the charge, the momentum, and the density in phase
space, integrated over the distribution in momentum space:
\begin{equation}
j^\alpha(x) = \int d^4 p \; e \, p^\alpha \, f(x,p) .
\label{e6}
\end{equation}
Beginning with this expression, one sees how integrals over
the distribution in momentum space naturally arise.
Also, notice that in $QED$, everything ---
$x$, $p$, $f(x,p)$, and $j_\alpha$ --- are all gauge invariant.

These elementary equations, (\ref{e3}) - (\ref{e6}), are all that
we need to solve for the effective action.  
Using the chain rule, the Boltzmann equation equals
\begin{equation}
\frac{d f}{d\tau} = 
\left( \frac{d x^\alpha}{d\tau} \frac{\partial}{\partial x^\alpha}
+ \frac{d p^\alpha}{d\tau} \frac{\partial}{\partial p^\alpha} \right) f  \; .
\end{equation}
Plugging in the equations of motion, 
\begin{equation}
\frac{d f}{d\tau} = p^\alpha \left(\frac{\partial}{\partial x^\alpha}
- e F_{\alpha \beta} \frac{\partial}{\partial p^\beta}\right) f = 0 \; ,
\end{equation}
which is known as the Vlasov equation.

Now expand the complete distribution function
about some inital value, $f^0$,
\begin{equation}
f = f^0 + f^1 + \ldots 
\end{equation}
The system must start out electrically neutral, with $f^0$ 
the same for positrons as for electrons, so initially there
is no current.  A current is induced by fluctuations,
$f^1$.  Solve the Boltzmann equation to first order
in $f^1$:
\begin{equation}
p\cdot\partial f^1 = e \, p^\alpha \, F_{\alpha \beta} 
\; \frac{\partial f^0}{\partial p_\beta}  \; .
\label{e6a}
\end{equation}
Without being too careful about what it means,
we rather sloppily introduce 
the nonlocal operator $1/p\cdot \partial$ to
solve (\ref{e6a}) for $f^1$, and obtain the induced current:
\begin{equation}
j^\alpha = \int d^4 p \;
e p^\alpha \frac{1}{p\cdot\partial} \; e p^\beta \;
F_{\beta \gamma}\;
\frac{\partial f^0}{\partial p_\gamma} \; .
\end{equation}
Now I follow Brandt, Frenkel, and Taylor \cite{frenkel}, 
and integrate $\partial/\partial p_\gamma$ by parts.
Remember that the momentum $p$ is just a parameter of
the initial (classical) distribution, so the field
strength tensor is completely independent of it.
It is then easy to integrate by parts,
\begin{equation}
j^\alpha = - e^2 \int d^4 p \;
\frac{\partial}{\partial p_\gamma}
\left(\frac{p^\alpha p^\beta}{p\cdot\partial} \right)\;
F_{\beta \gamma} \, f^0 \; .
\end{equation}
There are three terms from this derivative, which equal
\begin{equation}
j^\alpha = - e^2 \int d^4 p \;
\left( \frac{\delta^{\alpha \gamma}p^\beta}{p \cdot \partial}
- \partial^\gamma \frac{p^\alpha p^\beta }{(p \cdot \partial)^2} \right)
F_{\beta \gamma} \, f^0 \; .
\label{e7}
\end{equation}
Of the three terms, that 
$\sim \delta^{\beta \gamma}$ drops out, because
it is contracted with the antisymmetric field strength,
$F_{\beta \gamma}$.

Now, implicitly any current defines a lagrangian density
through the relation
\begin{equation}
j^\alpha = \frac{\delta {\cal L}_{Debye}}{\delta A_\alpha} \; .
\end{equation}
Thus we can need to find the ``Debye'' lagrangian, ${\cal L}_{Debye}$,
which generates the current of (\ref{e7}).  
This is easy: since ${\cal L}_{Debye}$ is gauge invariant,
it is natural to try to use the field
strength tensors, $\sim F_{\alpha \beta}$, since they are
automatically gauge invariant.  Without too much
effort one can see that the result is
\begin{equation}
{\cal L}_{Debye} = \frac{e^2}{2} \! \int \! \! d^4 p
\left( F_{\alpha \beta} 
\frac{p^\beta p^\gamma}{-(p \cdot \partial)^2}
F_{\alpha \gamma} \right) \, f^0(p) \; .
\label{e8}
\end{equation}

\section{Classical kinetic theory in $QCD$}

I defer any discussion of (\ref{e8})
to show that once one sets up the classical kinetic
theory in a properly gauge invariant fashion, 
the nonabelian Debye lagrangian follows immediately from the abelian.

So let me start by reviewing what nonabelian gauge invariance is.
The lagrangian for a massless quark coupled
to a $SU(N)$ gauge field looks just like that for $QED$, 
\begin{equation}
{\cal L} = \bar{\psi}\gamma^\alpha D_\alpha \psi
+ \frac{1}{2} tr(G_{\alpha \beta}^2) \, , 
\end{equation}
except that now the gauge potential is an $SU(N)$ matrix,
$A_\alpha = A^a_{\alpha} t^a$, where I normalize that
$SU(N)$ matrices $t^a$ as 
$tr(t^a t^b) = \delta^{a b}/2$, with
the indices $a,b = 1...(N^2-1)$.

Normally, textbooks present nonabelian gauge symmetry
as a direct generalization of the abelian symmetry, giving
the transformation 
of the gauge potential, etc.  This is a completely confusing
way of viewing nonabelian gauge invariance.  Instead of
considering the gauge potential, it is much easier
to concentrate on the covariant derivative, since
under a local gauge transformation, it transforms homogeneously:
\begin{equation}
D_\alpha = \partial_\alpha - i g A_\alpha 
\rightarrow \Omega^\dagger D_\alpha \Omega \; , \;
\Omega^\dagger \Omega = 1  \; .
\label{e8a}
\end{equation}
The transformation of the gauge potential $A_\alpha$ can be
worked out from this; its transformation is inhomogeneous
so inelegant.  Given (\ref{e8a}), however,
you don't need to worry about the transformation of
$A_\alpha$.  For instance, if the quark fields transforms
as in $QED$, like 
$\psi \rightarrow \Omega^\dagger \psi$, then obviously the 
quark part of the lagrangian is gauge invariant.

The field strenth is constructed from the commutator of two
covariant derivatives.  Since $D_\alpha$ transforms homogeneously,
so does its commutator:
\begin{equation}
G_{\alpha \beta} = \frac{1}{- i g} [D_\alpha , D_\beta]
\rightarrow \Omega^\dagger G_{\alpha \beta} \Omega \; . 
\label{e9}
\end{equation}
This is not obvious from the explicit expression for the field
strength tensor, 
\begin{equation}
G_{\alpha \beta} = \partial_\alpha A_\beta
- \partial_\beta A_\alpha
- i g [A_\alpha , A_\beta] \; ,
\end{equation}
and the transformation of $A_\alpha$.

The moral of the story is 
that nonabelian gauge invariance can be easily ensured 
just by sticking to covariant derivatives everywhere.
The derivation of the nonabelian Debye lagrangian below
provides an striking illustration of this.

I now turn to the classical equations of motion for a nonabelian
particle.  As usual the position and momentum are related as
$p^\alpha = d x^\alpha/d \tau$.
The generalization of the Lorentz force equation of
(\ref{e4}) is not trivial, though, since while
$x$ and $p$ are gauge invariant, in a nonabelian
theory (unlike the abelian case) the field strenth tensor 
$G_{\alpha \beta}$ is not.
To form a gauge invariant equation, it is necessary to
introduce a matrix valued charge, $Q$:
\begin{equation}
\frac{d p^\alpha}{d \tau} = 2 \; g \; tr(Q \, G_{\alpha \beta}) 
\; p^\beta  \; . 
\end{equation}
This is trivially gauge invariant if, like $G_{\alpha \beta}$,
$Q$ transforms homogeneously under a local gauge transformation:
\begin{equation}
Q \rightarrow \Omega^\dagger Q \Omega \; .
\end{equation}

We now have a different problem - what is the equation of
motion for $Q$?  This can basically be guessed 
from gauge and lorentz invariance.
We certainly want the equation of motion to be
gauge covariant, which is easily ensured by taking
$D_\alpha Q$.  This isn't quite right, however, because
there are then four, instead of one, equation of motion.
To get one equation, we contract $D_\alpha Q$ with the
obvious vector floating about, which is the momentum.
This gives us a result first obtained by S. Wong \cite{wong}:
\begin{equation}
\frac{d x^\alpha}{d \tau} \; D_\alpha Q = 0 \; .
\label{e9a}
\end{equation}

The phase space for a nonabelian particle is now $x$, $p$, and
$Q$; thus the single particle density is in principle
a function of all three,
$f(x,p,Q)$.  The classical current is again the product of
the charge, $g$ times $Q$, the momentum, and the phase
space density:
\begin{equation}
j^\alpha(x) = \int d^4 p \; \int dQ 
\; g \, Q \, \, p^\alpha \, f(x,p,Q) \; .
\label{e9b}
\end{equation}

I note that the equation of motion for the gauge field is
\begin{equation}
D_\alpha G^{\alpha \beta} = j^\beta \; .
\end{equation}
Since $G^{\alpha \beta}$ is a commutator of $D$'s, 
$D_\alpha D_\beta G^{\alpha \beta} = 0$, so
the current is covariantly conserved, 
\begin{equation}
D_\alpha j^\alpha = 0  \; .
\end{equation}
This is only true if $Q$ satisfies Wong's equation.

Taking these equations for granted, solving for the
nonabelian Debye action is no harder than for the abelian.
The collisionless Boltzmann equation. is unchanged:
\begin{equation}
\frac{d }{d \tau}\, f(x,p,Q) = 0 \; .
\end{equation}
The basic point is that because $x$, $p$, and $f(x,p,Q)$ are 
gauge invariant, while 
$Q$, $j^\alpha$, and the equations of motion are 
gauge covariant, then the resulting 
${\cal L}_{Debye}$ must be gauge invariant.

Now applying the chain rule to Boltzmann's equation 
gives a Vlasov equation with three terms:
\begin{equation}
\frac{d }{d \tau}\, f = 
p^\alpha \left( \frac{\partial}{\partial x^\alpha}
- 2 g \, tr \left(Q G_{\alpha \beta}\right)
\frac{\partial}{\partial p^\beta} 
+ 2 g \, tr\left( [A_\alpha,Q] 
\frac{\partial}{\partial Q}\right) \right)
f = 0 \; .
\end{equation}
The last term is from color precession of $Q$, and
is where all of the complications of the nonabelian
theory reside.

Suppose we only wish to compute to lowest order in
the gauge potential, however; this is linear in $A_\alpha$
for the current, or quadratic in $A_\alpha$ for the lagrangian.
Then the nonabelian theory is no more complicated than
the abelian.  Expand the distribution function about
some initial value, where the initial distribution is
assumed to depend only upon momentum:
\begin{equation}
f(x,p,Q) = f^0(p) + f^1(x,p,Q) + \ldots \; .
\end{equation}
Since $f_0$ is colorless, to lowest order
in $A$ we can drop the term for color precession in Boltzmann's
equation.  In this case the equations abelianize.
The {\it only} feature of the $Q$'s which is needed are the
Casimir's:
\begin{equation}
\int dQ \; Q^a Q^b = C \, \delta^{a b} \; .
\label{e10}
\end{equation}
For gluons in the adjoint representation, 
$C = N$, while for quarks in the fundamental representation,
$C = \frac{1}{2}$.

Thus to $\sim A^2$, the nonabelian debye action is a sum over
abelian actions:
\begin{equation}
{\cal L}^{(2)}_{Debye} = g^2 \! \int \! \! d^4 p \;
tr \left( F_{\alpha \beta} 
\frac{p^\beta p^\gamma}{-(p \cdot \partial)^2}
F_{\alpha \gamma} \right) \, \sum_r C_r f_r^0(p) \; .
\label{e11}
\end{equation}
Here the field strength is only the abelian part, linear in
$A_\alpha$: $F_{\alpha \beta} = \partial_\alpha A_\beta
- \partial_\beta A_\alpha$.  

It is then easy to make ${\cal L}_{Debye}^{(2)}$ gauge invariant:
replace the abelian part of the field strength by the
complete field strength, and replace the ordinary by the 
covariant derivative, to obtain:
\begin{equation}
{\cal L}_{Debye} = g^2 \! \int \! \! d^4 p
\; tr \left( G_{\alpha \beta} 
\frac{p^\beta p^\gamma}{-(p \cdot D)^2}
G_{\alpha \gamma} \right) \, \sum_r C_r f^0_r(p) \; .
\label{e12}
\end{equation}
The sum is over the representations of all charged fields.
Since everything transforms homogeneously under a local
gauge transformation, it is obviously gauge invariant.
The crucial question is whether the generalization is
unique.  On this point I have to defer to the literature
\cite{htl}-\cite{lebellac}.

In hot $QED$, 
(\ref{e8}) was first derived almost 40 years ago
by Silin \cite{silin}, 
using classical kinetic theory.
For an abelian theory, this is the complete result; 
${\cal L}_{Debye}$ is quadratic in the gauge potential,
and only contributes to the photon self energy, which is
how Silin wrote it.
In constrast, in a nonabelian theory, 
because of the $A_\alpha$ which lurks in the covariant
derivative in
$1/(p \cdot D)$, there are terms of arbitrary order in
the gauge potential.  This means that ${\cal L}_{Debye}$
contributes not just to the gluon self energy, but 
to couplings between three, four, or any number of gluons.
As a practical matter, this complicates using
the nonabelian Debye lagrangian in real complications,
although methods have been developed to deal with this \cite{htl}.

The nonabelian Debye lagrangian was first derived
by Taylor and Wong \cite{taylorwong}.
There are several, equivalent ways of writing it;
that in (\ref{e8}) is handy because it is manifestly
gauge invariant \cite{bp}.

\section{Thermal distribution}

To understand the significance of the nonabelian Debye
lagrangian, consider a thermal distribution
at a temperature $T$:
\begin{equation}
f^0 = \delta(p^2) \times
\frac{1}{e^{|p^0|/T}\mp1} \; ,
\end{equation}
where the $-$ is for the Bose-Einstein
distribution of gluons, and the $+$ for the Fermi-Dirac distribution
of quarks.
In this case it is easy to compute the nonabelian Debye lagrangian.
The integral over the magnitude of $p$ generates a factor
$\sim T^2$, and leaves an angular integral:
\begin{equation}
{\cal L}_{Debye}
= \frac{3}{2} \, m^2_g
\int \frac{d\hat{p}}{4 \pi} \; 
tr \left( G_{\alpha \beta} 
\frac{p^\beta p^\gamma}{-(p \cdot D)^2}
G_{\alpha \gamma} \right)  \; .
\label{ey}
\end{equation}
I redefine $p \rightarrow (i,\hat{p})$,
with $\hat{p}$ a unit spatial vector,
$\hat{p}^2=1$.  $m_g$ is the gluon Debye ``mass'',
\begin{equation}
m^2_g = \left( N + \frac{N_f}{2} \right) 
\frac{g^2 T^2}{9}  \; .
\end{equation}

There is also a Debye lagrangian for quarks:
\begin{equation}
{\cal L}^q_{Debye}
= m^2_q \int \frac{d\hat{p}}{4 \pi} \; 
\overline{\psi} \left( \frac{p \cdot \gamma }{p \cdot D} \right) \psi \; ,
\label{e13a}
\end{equation}
where $m_q=$ is the quark Debye ``mass'', 
\begin{equation}
m^2_q = \frac{N^2 - 1}{2 N} \; \frac{g^2 T^2}{8} \; .
\end{equation}
The quark Debye lagrangian is chirally symmetric, like
the original gauge interaction.

We can now understand the applicability of the Debye
lagrangians.  
For a massless gas at nonzero
temperature, the typical momentum is of order the
temperature, which is termed ``hard'' \cite{htl}.  
The nonabelian Debye lagrangian introduces ``soft'' momenta, 
on the order of the Debye masses, $m_g \sim m_q \sim gT$.
To understand the relative importance of fields, we
use power counting, taking the gauge
field $A_\alpha \sim T$.  Thus for hard momenta, the original
lagrangian ${\cal L} \sim T^4$, while the nonabelian 
Debye term is just part of the perturbative corrections,
${\cal L}_{Debye} \sim g^2 T^4$.  In contrast, for soft momenta,
both the original and the nonabelian Debye lagrangians
are of the same order, ${\cal L} \sim {\cal L}_{Debye}
\sim g^2 T^4$.  Thus for soft momenta, since
${\cal L}_{Debye}$ is as big as ${\cal L}$,
it is necessary to include both in an effective lagrangian.

Diagramatically, one can show that the nonabelian Debye
lagrangian only receives contributions from
one loop diagrams in which the loop momentum are hard.
This suggests the term ``hard thermal loops'', which is
commonly used in the literature.  In this talk I
adopt instead the more generic phrase of nonabelian
Debye lagrangian, which sounds less technical.
I emphasize that while the physics subsumes 
Debye screening, there is much more going on than just that.

Indeed, consider the textbook example of hot $QED$.
While the photon self energy is treated in the
textbooks \cite{silin}, that of hot fermions is not.
The Debye lagrangian for fermions in hot $QED$ is very
similar to that of quarks in hot $QCD$, (\ref{e13a});
except for a change in the Debye quark mass, $m_q$,
the only other change is to use the abelian covariant
derivative.  As noted, in hot $QED$ the abelian
Debye lagrangian just contributes to the photon self
energy, nothing more.  But the fermion Debye lagrangian
is as complicated in hot $QED$ as in hot $QCD$:
the nonlocal factor
of $1/p\cdot D$ generates not just a contribution to
the fermion self energy, but as well
an infinite series of couplings between a fermion anti-fermion
pair and any number of photons.  Thus the textbook treatment of
hot $QED$ is seriously incomplete.
There is a good reason for this: 
the fermion Debye lagrangian does not appear to be easily
derivable by the standard form of classical kinetic theory.
It has been derived in perturbation theory, and from the 
semiclassical kinetic
theory of Blaizot and Iancu \cite{htl};
perhaps it could be derived using
the approach outlined in sec. 6.

One novel aspect of the Debye lagrangians  
is that they are nonlocal, because of
the the factors of $1/p\cdot \partial$.  In a thermal distribution,
this gives rise to discontinuities for spacelike
momenta, which can be understood as a relativistic
generalization of Landau damping.  

There is a more general lesson.  
At zero temperature, a standard
assumption in constructing any effective lagrangian
is that all terms must be local.  The nonabelian Debye lagrangian
shows that this is no longer true at nonzero temperature,
although the nonlocality which enters is of an
extremely specific form.

The nonlocality can be made to (apparently) disappear
by choosing the gauge $p \cdot A = 0$.
In that case, the entire
lagrangian collapses to a mass term for the gauge field:
\begin{equation}
{\cal L}_{Debye} = g^2 \! \int \! \! d^4 p
\; tr \left( A_{\alpha}^2 \right) \, \sum_r C_r f^0_r(p) \; .
\label{e13}
\end{equation}
This form was first noted by Frenkel and Taylor \cite{htl}, \cite{frenkel}.  
Elmfors and Hansson have shown how it can be used to provide
a direct functional derivation of ${\cal L}_{Debye}$ 
from the background field method \cite{elm}.  Similarly, in this
gauge the quark Debye lagrangian also collapses to just a self
energy.  For a general distribution this gauge choice 
can only be imposed after generalizing
the gauge potential to be a function not just of spacetime,
but also of the momentum of the initial distribution, $p$.

\section{Tsunami wave distribution}

In the above I have made no reference
to the initial distribution, $f^0$.  It may be thermal,
but need not be; all that is necessary is that
the change from the initial distribution is
small, $f^1 \ll f^0$.  
For a thermal distribution, a small perturbation
will automatically return to thermal, but this
is certainly not true for an arbitrary initial
distribution.  

For a general distribution, the nonabelian Debye
lagrangian is presumably valid 
not just at weak coupling, but only for small
times; at long times, what $f^0$ evolves into
depends on the detailed dynamics.  I now apply these results
to another distribution, admittedly cooked up.

In cascade models of heavy ion collisions, the two colliding
nuclei are modeled by two pancakes, where in each pancake, all
partons move in lockstep with the same momentum.  Now 
take one pancake away, and enlarge the other pancake until
it fills up space uniformly.

This leads to the ``tsunami wave'' problem: at time $t=0$,
assume that one has a large and spatially
constant density of particles, all
moving together with the same momentum.  Into what state
does this evolve at infinite time?

The particles must be bosons, since I have assumed that the
initial density is high.  If the particles have nonzero mass,
then one can transform into their rest frame, in which case
the particles are just a condensate (of some sort) at
zero momentum.  So assume that they are massless, and move
on the light cone.  

The initial state is a system with nonzero energy and momentum
density.  Thus one natural guess for the state at infinite
time is that of a boosted thermal distribution, since that
also has nonzero energy and momentum density.
That the system thermalizes is not obvious; assuming that it
does, the relevant question then is, over what time scales?

This is a very difficult question which I could not attempt
to solve analytically.  For small times, however, a 
perturbative analysis should be a reasonable approximation.
I define the tsunami wave distribution as:
\begin{equation}
f^{0}(p) \sim \rho_0 \; \delta(p^2) \; |p_0| \;
\delta^3(\vec{p}-\vec{p_0}) \; .
\label{e14}
\end{equation}
Here $\rho_0$ is a parameter proportional to the density,
and $p_0 = (i |p_0|, \vec{p_0})$, $p_0^2 = 0$, is the momentum
of the particles in the tsunami wave. 

We now have a problem in which there is a preferred four vector,
$p_0$.  For this problem the obvious choice of gauge is
$p_0 \cdot A = 0$.  Consider the gluon self energy:
it is a function of the four momentum, $k$, and also of
this vector $p_0$.  In general four functions enter into
the self energy:
\begin{equation}
\Pi^{\alpha \beta}(k)
= \Pi_t \; \delta^{\alpha \beta}
+ \Pi_\ell \; k^\alpha k^\beta
+ \Pi_3 (k^\alpha p_0^\beta + p^\alpha_0 k^\beta)
+ \Pi_4 \; p_0^\alpha p_0^\beta  \; .
\label{e15}
\end{equation}
This is analogous to the situation at nonzero temperature,
where the rest frame of the thermal bath provides a preferred
four vector.  For the thermal case, the preferred vector is
timelike; here it is null.  Also as at nonzero temperature,
the four functions in $\Pi^{\alpha \beta}$ are related by
a Ward identity, but that doesn't matter here.

We can then read off the result from (\ref{e13}).
Ignoring inessential constants,
\begin{equation}
\Pi_t \sim + g^2 N \rho_0 \;\;\; , \;\;\;
\Pi_\ell = \Pi_3 = \Pi_4 = 0 \; \; .
\label{e17}
\end{equation}

This is not the whole story, however; it is still necessary
to work out the effective propagator, including this self
energy.  This is a straightforward if tedious exericse.
In $p_0 \cdot A = 0$ gauge, the result is
\begin{equation}
\Delta^{\alpha \beta}
= \frac{1}{k^2 + \Pi_t} \left( \delta^{\alpha \beta}
- \frac{p_0^\alpha k^\beta + k^\alpha p_0^\beta}{p_0 \cdot k}
- \frac{k^2}{k^2 - \Pi_\ell} (\Pi_t + \Pi_\ell)
\frac{p_0^\alpha p_0^\beta}{(p_0 \cdot k)^2} \right) \; .
\label{e16}
\end{equation}

Thus we see that in a tsunami wave, the two, spatially
transverse modes which one expects for a spin-one field
are screened.  If $\Pi_\ell$ were nonzero, it would
represent a collective mode, analogous to the plasmon
of a thermal distribution.  Because $\Pi_\ell = 0$,
there is no plasmon for this type of tsunami wave.

Thus at short times, interactions between gluons in a dense
tsunami act to screen the usual transverse modes of the gluon.
What happens at longer times is an open question.  This
can be studied numerically in a scalar theory in the limit
of a large number of components \cite{boy}.

\section{Where does $Q$ come from?}

While classical kinetic theory is often used in nonrelativistic
systems, its use in a nonabelian gauge theory is unfamiliar.
What are the original degrees of freedom in the gauge theory
which conspire, in the classical limit, to generate the
nonabelian charge $Q$?

Let me start with something familiar, the lagrangian
for a charged scalar field, $\phi$, in the presence of a
background gauge field:
\begin{equation}
{\cal L} = \phi^\dagger (- D^2) \phi \; .
\end{equation}
Integrating over $\phi$ gives the effective action
\begin{equation}
{\cal S}_{eff} =  tr \; log \; \left( - D^2 \right) \; .
\label{eff}
\end{equation}

Now I use the usual trick of Feynman and Schwinger, to
turn this ${\cal S}_{eff}$ into the path integral for
a particle.  First introduce a parameter, $\tau$,
which will become like a proper time:
\begin{equation}
{\cal S}_{eff} = \int^\infty_0 \frac{d \tau}{\tau}
\; tr\left(e^{- \frac{1}{2} \tau D^2}\right) \; .
\end{equation}

Going to momentum space,
\begin{equation}
{\cal S}_{eff} = \int^\infty_0 \frac{d \tau}{\tau}
\int \frac{d^4 p}{(2 \pi)^4} \;
tr\left(e^{- \frac{1}{2} \tau (p- g A)^2}\right) \; .
\end{equation}
The remaining trace is only for the $SU(N)$ matrix $A_\alpha$.
This is then converting into a sum over paths:
\begin{equation}
{\cal S}_{eff} \sim \int^\infty_0 \frac{d \tau}{\tau}
 \int {\cal D}x {\cal D}p 
\; tr \left( e^{- \int^\tau_0 d\tau ' \; {\cal L}(p)} \right) \; ,
\end{equation}
\begin{equation}
{\cal L}(p) = - i p\cdot \dot{x}
+ \frac{1}{2} (p - g A)^2 \; .
\label{e17a}
\end{equation}
where $\dot{x}^\alpha = dx^\alpha/d\tau'$.
The integral over the momentum is trivial,
\begin{equation}
{\cal S}_{eff}
\sim  \int^\infty_0 \frac{d \tau}{\tau}
\int {\cal D} x \; tr \left( e^{- \int^\tau_0 d\tau ' 
\; {\cal L}_A} \right) \; ,
\label{e18}
\end{equation}
\begin{equation}
{\cal L}_A =  \frac{\dot{x}^2}{2} - i g A\cdot \dot{x} \; ,
\label{e19}
\end{equation}
subject to the boundary conditions $x^\alpha(\tau)= + x^\alpha(0)$

This form of the one loop effective action has been studied
by many authors \cite{strassler}.  
This representation has been shown to be 
a powerful way of reorganizing perturbation theory.  
From (\ref{e18}), this is just a matter of
expanding the exponent in powers of the gauge field.

In a more general context, however, ({\ref{e18}) is
incomplete.  In (\ref{e19}), the ``lagrangian'' ${\cal L}$
is not really that at all: it is a matrix, as in (\ref{e18})
there is still a color trace left to do.
What we want instead is a form where the lagrangian is 
a true scalar in color space.

This can be done by replacing 
the Wilson line by an integral over worldline
fermions, $\lambda(\tau)$ \cite{prev,dhoker}:
\begin{equation}
tr\; {\cal P} \; e^{ i g \int^\tau_0 A \cdot dx}
= \left( \frac{\pi}{\tau}\right)^N 
\sum_{k=1}^{N} 
\int {\cal D}\lambda {\cal D}\lambda^\dagger
\; e^{i k ( \lambda^\dagger \lambda + N/2 - 1)}
\; e^{- \int^\tau_0 d\tau ' {\cal L}_\lambda} \; ,
\label{e19a}
\end{equation}
\begin{equation}
{\cal L}_\lambda
= \lambda^\dagger \left( \dot{x} \cdot D \right) \lambda \; .
\label{e20}
\end{equation}
The $\lambda(\tau)$ lie in the fundamental representation of
$SU(N)$ color, and satisfy antiperiodic boundary conditions,
$\psi(\tau) = - \psi(0)$.  Path ordering is denoted by
${\cal P}$, and is automatic in the path integral.

It has been known for some time that worldline fermions
are required to describe finite dimensional representations
of nonabelian charge \cite{prev}.  
To understand (\ref{e19a}), notice that in the absence 
of a gauge field, the propagator for a fermion 
in one dimension is a step function: the solution to
$\partial_\tau \Delta_\lambda = \delta(\tau)$ is 
$\Delta_\lambda \sim \theta(\tau - \tau ')$.  
Then, since propagation is
in one dimension, the complete propagator is naturally
an exponential, where the step function provides the path ordering.

The complete form in (\ref{e19a}), including
the sum over $k$, was derived by D'Hoker and Gagn\'{e},
eq. (5.2) of \cite{dhoker}.  The sum over $k$ is
required in order to project upon
states with occupation number one.

The integral over worldline fermions then replaces the color
sum, and gives us a true lagrangian.
We can then use this
to go back, reintroduce the momentum conjugate to the position
$x$ (this is not necessary for the worldline fermions,
since $\psi^\dagger$ is already conjugate to $\psi$), to obtain
\begin{equation}
{\cal S}_{eff} 
\sim  \int^\infty_0 \frac{d \tau}{\tau}
\int {\cal D} x {\cal D}p {\cal D}\lambda
{\cal D}\lambda^\dagger  \;   
\sum_{k=1}^{N} \; e^{i k ( \lambda^\dagger \lambda + N/2 - 1)} \;
\left( \frac{\pi}{\tau}\right)^N  \;
e^{- \int^\tau_0 d\tau ' \; {\cal L}}  \; ,
\label{e21}
\end{equation}
\begin{equation}
{\cal L} = - i p \cdot \dot{x} - \lambda^\dagger \dot{\lambda}
+ \frac{1}{2}\left(p - 2 g \, tr(Q A) \right)^2 \; .
\label{e22}
\end{equation}
The nonabelian charge $Q$ is just
\begin{equation}
Q^a = \lambda^\dagger t^a \lambda \; .
\label{e23}
\end{equation}
The lagrangian ${\cal L}$ of (\ref{e22}) is similar
to the original ${\cal L}(p)$ of (\ref{e17a}), but
now the color trace downstairs is replaced by
an integral over the worldline fermions.  Moreover,
the nonabelian charge now enters in a most natural
manner, as a bilinear in the worldline fermions.

It is easy to check that ${\cal L}$ gives the correct
equations of motion for a classical, nonabelian particle.
For example, evidently the equation of
motion for the worldline fermion, $\dot{x} \cdot D \lambda = 0$,
gives Wong's equation for $Q$, (\ref{e9a}).  

A lagrangian very similar to ${\cal L}$ was proposed by
Brandt, Frenkel, and Taylor \cite{frenkel} to
generate the nonabelian Debye lagrangian.  
(They used wordline scalars instead of fermions, but
this difference doesn't matter in the classical limit \cite{prev}.)
My contribution is to note 
that the introduction of the worldline fermions
is not just a trick to get the correct equations of motion,
but is a systematic approximation to the correct effective action.

Given this lagrangian, we then adopt a classical approximation.
Instead of an integral over the nonabelian charge $Q$, one
should integrate over the worldline fermions.  Since all we
needed before was the Casimir of the representation,
this doesn't matter in the classical limit.

Other approximation schemes have been developed to
analyze worldline path integrals at nonzero temperature.  
For instance, one can sum over paths which wind around in the imaginary
time direction \cite{mckeon}.  The resulting expressions
are not especially simple, though; classical
kinetic theory appears to be more useful.

However, this begs the question of how the collisionless
Boltzmann equation arises.  I conclude with a suggestion.
In the absence of interactions, at nonzero temperature 
the propagator is
directly proportional to the statistical distribution function:
\begin{equation}
tr \left(\frac{1}{-\partial^2} \right)
\sim n(p) \; .
\end{equation}
In the presence of a background gauge field, 
the propagator can be written as a sum over paths:
\begin{equation}
tr \left( \frac{1}{-D^2} \right)
\sim  \int^\infty_0  d \tau 
\int {\cal D} x {\cal D}p {\cal D}\lambda
{\cal D}\lambda^\dagger  \;   
\sum_{k=1}^{N} \; e^{i k ( \lambda^\dagger \lambda + N/2 - 1)} \;
\left( \frac{\pi}{\tau}\right)^N  \;
e^{- \int^\tau_0 d\tau ' \; {\cal L}}  \; .
\end{equation}
From this, define the ``single particle'' density, $f$, as
\begin{equation}
tr \left( \frac{1}{-D^2} \right)
\sim  \int^\infty_0 d \tau 
\int {\cal D} x {\cal D}p {\cal D}\lambda
{\cal D}\lambda^\dagger  \;  f(x,p,Q) \; ,
\end{equation}
Perhaps in the classical approximation, the dominant term
is that where $f$ is stationary with respect to $\tau$;
this would then give the collisionless Boltzmann equation.  
The expression for the current, (\ref{e9b}), follows
by differentiation with respect to the gauge field.
At two loop order, the effective action is not given
by (\ref{eff}), and is how collisions enter into the
Boltzmann equation.  

\section{Acknowledgements}

I thank M. H. G. Tytgat for discussions, and especially
for bringing (\ref{e19a}) to my attention.

{}  
\end{document}